\documentstyle[preprint,aps,pre,epsf,amsbsy,amssymb]{revtex}
\begin{document}
\draft
\newcommand{\ve}[1]{\boldsymbol{#1}}

\title{Catalytic CO Oxidation on Nanoscale Pt Facets: Effect of Inter-Facet CO
Diffusion on Bifurcation and Fluctuation Behavior} \author{N.~Pavlenko$^1$\cite{new_addr},
J.W.~Evans$^{2,3}$, Da-Jiang Liu$^2$, and R.~Imbihl$^1$} 
\address{$^1$Institut
f\"{u}r Physikalische Chemie und Elektrochemie, Universit\"{a}t Hannover,
Callinstr.~3-3a, D-30167, Hannover, Germany\\ $^2$Ames Laboratory and
$^3$Department of Mathematics, Iowa State University, Ames, Iowa 50011}
\date{\today} 
\maketitle 
\begin{abstract} 
We present lattice-gas modeling of the
steady-state behavior in CO oxidation on the facets of nanoscale metal clusters,
with coupling via inter-facet CO diffusion. The model incorporates the key
aspects of reaction process, such as rapid CO mobility within each facet, and
strong nearest-neighbor repulsion between adsorbed O. The former justifies our
use a "hybrid" simulation approach treating the CO coverage as a mean-field
parameter. For an isolated facet, there is one bistable region where 
the system can exist in either a reactive
state (with high oxygen coverage) or a (nearly CO-poisoned)
inactive state. Diffusion between two facets is shown to induce complex
multistability in the steady states of the system. The bifurcation diagram
exhibits two regions with bistabilities due to the difference between adsorption
properties of the facets. We explore the role of enhanced fluctuations in the
proximity of a cusp bifurcation point associated with one facet in producing
transitions between stable states on that facet, as well as their influence on
fluctuations on the other facet. The results are expected to shed more light on
the reaction kinetics for supported catalysts.
\end{abstract}


\section{INTRODUCTION}

A central problem in the modeling of catalytic surface reactions is to bridge the 
gap between the studies of reactions on ideal homogeneous surfaces of
macroscopic size and real catalytic processes on supported catalysts. The latter
are composed of small nm-sized metallic clusters on an
inert support material. The surface of these clusters contains facets of different 
orientations, and hence differing reactivities, which are coupled, e.g., through
diffusion of mobile adsorbates. Due to
this coupling effect, and since fluctuations in such small size facets can become 
dominant, the behavior of these systems can differ strongly from that predicted
by mean-field rate laws on ideal single-crystal surface planes. An understanding
of the effects that can arise is crucial not only for
the modeling of real catalysis, but also for the design of nanostructured catalysts.

For catalytic CO oxidation, by far the majority of experimental studies have been 
performed on extended single-crystal substrates. Mean-field equations, ignoring
spatial correlations and ordering in surface adlayers, have been successfully
applied to predict a wide variety of surface phenomena, including nonlinear
dynamics and spatial pattern formation \cite{kapral}. More sophisticated
lattice-gas models can be utilized to describe the effect of adspecies 
interactions, and "hybrid" formulations can treat directly and efficiently rapid
surface diffusion of some adsorbates such as CO (the rate of CO diffusion is of
many orders of magnitude higher than the rates of adsorption,
desorption and O diffusion). A recently developed canonical hybrid lattice-gas 
model of this type for CO-oxidation incorporates a random distribution of CO,
and strong nearest-neighbor O repulsion \cite{tamarro,james,zhdanov1}.

In-situ experimental studies of reactions on supported catalysts are still beyond 
the imaging capabilities of current microscopes. However, for a reasonable model
system as far as the dimension of the facets and the coupling of different
orientations are concerned, one can consider reactions on a field emitter tip.
The surface of such a tip can be imaged with field electron/field ion
microscopy with nearly nanometer/atomic resolution. In a series of recent 
experiments, the spatio-temporal dynamics of the fluctuations which arise in 
catalytic CO oxidation on a Pt tip have been studied in some detail \cite{fluct}.

Conventional or hybrid lattice-gas modeling is ideally suited to analyze both 
kinetics and fluctuations in nanoscale systems \cite{fluct,zhdanov2}. For example,
recent modeling of reaction kinetics on the facets coupled by reactant diffusion
reveals shifts in the reaction window for bistable systems, and shifts in the
oscillation window for systems which incorporate additional feedback. The latter 
also indicates a possible transition from regular to chaotic oscillations due to
additional supply of reactants from adjacent facet \cite{zhdanov2}. Lattice-gas
modeling has also been applied to analyze fluctuations in catalytic CO oxidation
within a small isolated facet containing hundreds to thousands of adsorption sites
\cite{fluct}. What has been neglected in these studies \cite{fluct} is the coupling
between adjacent facets on the Pt field-emitter tip. Experimentally, there is some
justification for this, since one observes that fluctuations on different facets were
usually largely uncorrelated (while a high degree of spatial correlations exists only
for fluctuations on a single facet). One can in fact explain this
behavior noting that CO diffusion is very fast within a facet but inhibited by 
structural heterogeneities (such as atomic steps) at the periphery of the facet.
However, there is some coupling between facets, as the kinetics is described by
a global bifurcation diagram, and there must be some correlation between the
fluctuations.

Thus, to shed more light on reaction behavior for coupled nanoscale-facets, we 
focus in this work on incorporation of inter-facet diffusion into the
above-mentioned canonical model describing key aspects of CO oxidation. We find
that the diffusive coupling between two adjacent facets with different
adsorption properties results in the appearance of complex multistability of the 
steady states (which is fully characterized exploiting a novel simulation
approach). Consequently, the bifurcation diagram exhibits two regions with
bistabilities instead of single bistable region observed
in the case of isolated facet. We show that typically the different facets have 
different fluctuation characteristics. However, in some cases, the
diffusion-induced coupling can change radically the
fluctuation behavior of whole system, producing a transition between stable 
states on one facet induced by enhanced fluctuations on the other facet.

\section{THE TWO-FACET REACTION MODEL AND ITS ANALYSIS}

\subsection{Model Specification}

The lattice-gas model employed for our studies is based on Langmuir-Hinshelwood 
mechanism for catalytic CO oxidation. The two "adjacent" small facets labeled
$i=1,2$ represented by two rectangular grids of $N_i=L_0 \times L_i$ adsorption
sites. One could physically connect the facets, e.g., at
a common edge of length $L_0$, which would introduce spatial inhomogeneity within 
each facet.
However, we adopt a simplified treatment invoking periodic boundary conditions. We
incorporate the following steps into a "hybrid" model \cite{tamarro,james}, adapted
here for the two-facet case:\\
(i) CO(gas) adsorbs onto single empty sites at rate $p_{\rm CO}$ per site on both
facets, and desorbs from at rate $d$. CO(ads) can hop very rapidly to other empty
sites within each facet. Since we consider below the regime of infinitely mobile
CO(ads) within the facets, and neglect interactions between CO(ads) and other
CO(ads) and O(ads), the distribution of CO(ads) on sites not occupied by O(ads)
is random. Thus, in simulation procedure, we track only the total numbers
$N_{\rm CO}^1$ and $N_{\rm CO}^2$ of CO(ads) on both facets, whereas distribution of
O(ads) for each facet is described by a full lattice-gas approach.\\
(ii) CO(ads) can diffuse at a finite rate from each facet to empty sites 
on the adjacent facet. Physically, CO(ads) would hop across a common edge (of
length $L_0$) to adjacent empty sites at rate, say, $h'$. In our model with
periodic boundary conditions, this mass transport is mimicked by transport of
CO(ads) from facet $i$ to $j$ at rate (in molecules per unit time) $h_i
N_{\rm CO}^i$ times the probability that a site is empty on facet $j$, where 
$h_i=h'/L_i$.\\
(iii) O$_2$(gas) adsorbs dissociatively at the impingement rate $p_{{\rm O}_2}$ per
site (as described below), but the sticking coefficients for oxygen differ for each
facet: $s_{\rm O}^1 \neq s_{\rm O}^1$ so that the two facets exhibit a different
catalytic activity. To account for very strong NN O(ads)-O(ads) repulsions, we invoke
an "eight-site rule" \cite{8cite} wherein O$_2$(gas) adsorbs only at diagonal
nearest-neighbor (NN) empty sites, provided that the additional six sites adjacent to
these are not occupied by O(ads).
Also, O(ads) is immobile, and can not desorb, and thus never occupies adjacent
sites within each facet.\\
(iv) Adjacent pairs of CO(ads) and O(ads) within a facet reacts at rate $k$ to 
form CO$_2$(gas). 

As is the single-facet case, we set $p_{\rm CO}+p_{{\rm O}_2}=k=1$ and consider system
behavior as a function of $p_{\rm CO}$ with $p_{{\rm O}_2}=1-p_{\rm CO}$. In contrast
to the simplified ZGB (Ziff-Gulari-Barshad) model \cite{zgb}, where CO(ads) are immobile, 
the main feature of the above model when applied to single isolated facet is the appearance of strong bistability:
over a significant range of $p_{\rm CO}$ and (smaller) $d$ values, there are both stable 
reactive state (with high O coverage $\Theta_{\rm O}$) and 
stable inactive state (with the surface nearly poisoned by CO).

\subsection{(Standard) Constant-Partial-Pressure and (Refined) Constant-Coverage 
Simulations}

Standard "constant (partial) pressure" kinetic Monte Carlo (KMC) simulations simply
implement all processes (adsorption, desorption, reaction, diffusion) with the 
appropriate relative rates. They can be readily implemented for single and
coupled facet models, and enable analysis of
stable steady states as well as kinetics. To also analyze the non-trivial 
{\it unstable} steady states existing in bistable (or multistable) regions, a
constant-coverage algorithm \cite{ziff_bros} has been applied previously for large
single facet systems. In the adsorption part of the algorithm, instead of
attempting to deposit with the appropriate relative rates, these simulations 
specify a target CO coverage $\Theta_{\rm CO}$, and deposit CO (O) for actual CO
coverages below (above) this target. Then the asymptotic $p_{\rm CO}$ value
(corresponding to given $\Theta_{\rm CO}$) is found as a fraction of
attempts to deposit CO. In our case of two communicating facets with different O
sticking properties, the above constant-coverage approach is unsuitable: fixing the CO
coverage on both facets corresponds to different values of $p_{\rm CO}$ (which is
unphysical). To resolve this problem, in this work, we develop a combined
constant-pressure \& constant-coverage algorithm. Analysis using this novel simulation
procedure is performed in {\it two steps}.

{\it First}, we fix a target CO coverage for the first facet $\Theta_{\rm CO}^1$, but
use the constant-pressure algorithm for the second facet. Fixing $\Theta_{\rm CO}^1$ 
during this step allows us to probe both stable states together with unstable state in
the bistability region on the first facet, but the CO coverage on the
second facet $\Theta_{\rm CO}^2$ relaxes with time to a stable steady-state value
corresponding to the rate $p_{\rm CO}$ (which is equal for both facets). Because
of this feature, we can not probe the unstable state on the second facet in the
bistable region (the same limitation as in constant-pressure simulations).

{\it Second}, to probe the full set of states on the second facet (including the 
unstable state), we invert the above procedure, prescribing the coverage $\Theta_{\rm
CO}^2$ and performing standard constant-pressure simulations on the first facet. 
This gives us all states for the second facet, but just the stable states for
the first one (matching results from above). Thus, whereas the first step gives us
the full picture for the first facet and incomplete data for the second one, after 
performing both steps, we can obtain the full picture (i.e., the complete
bifurcation diagram) for the whole system.

\subsection{Analytic Pair-Approximation Treatment}

Our novel simulation procedure extends conventional KMC approaches in order 
to obtain complete steady state behavior. It is thus natural to compare results
with those obtained from analysis of (approximate) rate equations for our
two-facet CO oxidation (which automatically provide complete steady state
information). For convenience, we denote adsorbed CO by $A$ and
adsorbed O by $B$, using standard notation for monomer-dimer $A+B_2$ reaction model. 
Following \cite{james}, we let $E$ denote empty sites, $Z=1-B$ denote sites not
occupied by
B, and indicate probabilities of various configurations by square parentheses.
Then, for CO and O coverages on the $i$th facet ($i=1,2$) of $\Theta_{\rm
CO}^i=[A_i]$ and $\Theta_{\rm CO}^i=[B_i]$, respectively, the rate equations
have the following form:
\begin{eqnarray}
&& \frac{d}{dt}[A_i]=p_{\rm CO}[E_i]-d[A_i]-4k[A_iB_i]-{h_i}J_i,
\label{rate_eq} \\
&& \frac{d}{dt}[B_i]=2p_{{\rm O}_2} s_{\rm O}^i
\left[
\begin{array}{cccc}
 & & Z & \\
 & Z & E & Z \\
 Z & E & Z &\\
 & Z & &
\end{array}
\right]_i-4k[A_iB_i], \quad (i=1,2). \nonumber
\end{eqnarray}                                 
The net diffusive CO flux $J_1=-J_2=[A_1E_2]-[A_2E_1]=[A_1(1-B_2)]-[A_2(1-B_1)]$
introduced in (\ref{rate_eq})
describes the CO transport between facets with the microscopic hop rate $h'$ related 
to the $h_i$ as described above. (In our analysis, we set $N_1=N_2$ corresponding to
equal size facets, and thus $h_1=h_2=h$.) The lack of spatial correlations
between facets further reduces the $J_i$ to $J_1=-J_2=[A_1](1-[B_2])-[A_2](1-[B_1])$.
The lack of NN O(ads), together ´with the random distribution of CO within each facet,
allows further reduction of (\ref{rate_eq}). For example, the reaction terms can be
exactly expressed in terms of the coverages \cite{james}:
$[A_iB_i]=[A_i][B_i]/(1-[B_i])$.

These results allow derivation of an exact relationship between $[A]$ and $[B]$ for
the steady states (${d}/{dt}[A_i]=0$) from the first equation in (\ref{rate_eq}): 
\begin{eqnarray} 
&&[A_1]=-p_{\rm CO} (1-[B_1])(h_1(1-[B_2])+b_1)/c \nonumber\\
&&[A_2]=-p_{\rm CO} (1-[B_2])(h_2(1-[B_1])+b_2)/c, \label{a_12} 
\end{eqnarray} 
where $c=h_1h_2(1-[B_1])(1-[B_2])-b_1b_2$; $b_1=p_{\rm
CO}+d+4k[B_2]/(1-[B_2])+h_2(1-[B_1])$
and $b_2=p_{\rm CO}+d+4k[B_1]/(1-[B_1])+h_1(1-[B_2])$. We next substitute the
resulting expressions (\ref{a_12}) for $[A_i]$
into the second equation (\ref{rate_eq}) and applying the pair (Kirkwood-type)
approximation \cite{tamarro,james} to decouple the eight-site correlation functions in
(\ref{rate_eq}) describing the "eight-site rule". The analysis of the steady states for
two adjacent facets (corresponding to the conditions ${d}/{dt}[A_i]=0$ and
${d}/{dt}[B_i]=0$) is then reduced to the solving the closed system of two high-order
polynomial equations with respect to $[B_i]$, and subsequent evaluation of $[A_i]$
from (\ref{a_12}).

\section{RESULTS FOR STEADY-STATE BIFURCATION BEHAVIOR}

\subsection{Case Study: $s_{\rm O}^1=0.2$ and $s_{\rm O}^2=1.0$.}

We now present a brief overview of new results from our novel KMC simulation procedure
for coupled $N_a \times N_2$ site facets, which are compared with pair-approximation predictions
using rate equations (\ref{rate_eq}). To demonstrate the effect of inter-facet
diffusion,
for $h=0.01$, we show in Fig.~1 the CO coverages vs. $p_{\rm CO}$ for $s_{\rm O}^1=0.2$
and $s_{\rm O}^2=1.0$. A very good agreement is found between results from KMC
simulation and the pair-approximation (\ref{rate_eq}). The latter produces only slight
deviation of bistable regions towards lower $p_{\rm CO}$. The difference between
$[A_i]$ and $[B_i]$ obtained using both approaches is the most significant in the reactive
state, since the high O coverage makes the pair-approximation for "eight-site-rule" less
accurate. In contrast to the case of isolated facets when $h=0$ (shown in the insets of
Fig.~1), a qualitatively new feature appears in $\Theta_{\rm CO}^1$ and $\Theta_{\rm
CO}^2$ behavior due to the inter-facet communication by CO hopping. Besides the
bistability region (\#1) for $0.113<p_{\rm CO}<0.1515$ on the first facet with $s_{\rm
O}^1=0.2$ (resembling closely that for $h=0$), an additional bistable region (\#2) at
$0.17<p_{\rm CO}<0.3614$ forms on the first facet covered predominantly by CO. This is
a response to the bistable behavior in this range of $p_{\rm CO}$ values occurring on the second
facet with $s_{\rm O}^2=1.0$. Analogously, an additional bistable region on the second
facet for $0.113<p_{\rm CO}<0.1515$ arises in the state with low CO coverage as a
response to a bistable behavior on the first facet.  

Fig.~2 shows the variation with $h$ of the bifurcation diagrams in the ($p_{\rm
CO}$, $d$)-plane. For low diffusion rate $h$, the phase diagram contains two overlapping
bistable regions (see Fig.~2(a)): the small region \#1 at lower $p_{\rm CO}$ 
terminating at cusp point ($p_c^1 \approx 0.1548$, $d_c^1 \approx 0.025$)
originates from the first facet
($s_{\rm O}^1=0.2$), whereas the larger region \#2 terminating at cusp point 
($p_c^2 \approx 0.409$, $d_c^2 \approx 0.051$) occurs due to the second facet.
Note that the behavior of the both bistable regions differs dramatically with varying
h. The larger region \#2 varies only weakly with $h$, whereas the smaller region \#1
first shrinks, and then disappears. The latter behavior can be understood as CO on the
first facet is supplied very rapidly to the second facet with lower CO coverage, thus
suppressing completely the occurrence of region \#1. 

\subsection{Dependence of Behavior on Sticking Probabilities}

In extending our analysis to general $s_{\rm O}$ values, we introduce three different 
coupling regimes: (i) $0<h<h_1$, weak-coupling regime, when two overlapping bistable
regions cross each other on the diagram ($p_{\rm CO}$, $d$); (ii) $h_1<h<h_2$, 
medium-coupling regime with one smaller bistability locating completely inside another
larger one; (iii) $h>h_2$, strong-coupling regime with only one common bistable region.
This classification will be valuable for the studies of fluctuation behavior below. The
values $h_1$ and $h_2$ separating these regimes are shown in Fig.~3(a) as the functions
of $s_{\rm O}^1$ (at $s_{\rm O}^2=1.0$). As $s_{\rm O}^1 \rightarrow 0$, obviously both
$h_1$ and $h_2 \rightarrow 0$ (since the first facet becomes an oxygen-free "sink"
for CO, and thus the bistability region \#1 disappears, see the inset of
Fig.~3(a)).

The variation of $h_1$ and $h_2$ with $s_{\rm O}^1$ shows three distinct branches
arising from competition between the effects of: (i) the inequality in oxygen
sticking, and (ii) finite inter-facet transport. The {\it first branch} ($s_{\rm
O}^1<0.65$) is formed due to the increased $s_{\rm O}^1$ value providing the
expansion of the first smaller bistable region \#1. As the difference in oxygen
sticking properties $|s_{\rm O}^1-s_{\rm O}^2|$ decreases, the {\it second branch}
($0.65<s_{\rm O}^1<1$) appears due
to a dominant role of restricted inter-facet transport. We comment in particular on the
case $s_{\rm O}^1=s_{\rm O}^2=1$ (identical facets) where $h_2 \approx 0.06 \neq 0$,
and where one might expect just one bistable region (so $h_2=0$). Indeed, the
"symmetric" steady-states for a single facet must apply to this system, but
restricted inter-facet transport produces two additional "unsymmetric" stable
steady-states (see Fig.~3(b), inset). The region of the existence of these two
states located inside the large bistable region \#2, is indicated on the 
bifurcation diagram in Fig.~3(b). For the {\it third branch} ($s_{\rm O}^1>1$),
the inequality in sticking again begins to be of primary importance producing the
increase in $h_1$ and $h_2$ shown in Fig.~3(a).

\section{RESULTS FOR FLUCTUATION BEHAVIOR}

Bearing in mind the classification of coupling regimes in Sec.IIIB, we turn now to the
analysis of the coverage fluctuations. The fluctuation amplitudes depend inversely on
system size, and become more significant approaching a cusp critical point (analogous to
fluctuation behavior near a critical point in equilibrium thermodynamics). Thus, they
can induce transitions between the stable steady states for systems which are small
enough, or in a close proximity to a cusp point \cite{fluct,fichthorn}. Here, we
consider two coupled facets, each with 30x30 lattice sites, and analyze fluctuation
effects both for different coupling regimes, and for different $s_{\rm O}^1$ values at
fixed $s_{\rm O}^2=1.0$. As already noted, the size and location of the small
bistability region \#1 depends strongly on the value of $|s_{\rm O}^1-s_{\rm
O}^2|$. We show below
that this factor, together with the coupling strength ($h$), strongly
influences fluctuation behavior. We consider separately two regimes of coupling 
strength.

\subsection{Weak Coupling Regime}

In Fig.~4, we plot the normalized amplitudes $F_j^i=\left[ N_i \langle \delta
(\Theta_j^i)^2\rangle / \left(\langle \Theta_j^i\rangle(1-\langle \Theta_j^i\rangle)
\right)\right]^{1/2}$
($j=$~CO, O) ($\langle \Theta_j^i\rangle$ denote the tome averages of $\Theta_j^i=N_j^i/N_i$ and 
$\delta \Theta_j^i=\Theta_j^i-\langle \Theta_j^i\rangle$) vs. $d$ on
both facets close to cusp point ($p_c^2$,$d_c^2$) terminating
the large bistability region \#2 for $s_{\rm O}^1=0.7$ and $h=0.005$ (in this case
$p_c^2 \approx 0.4096$). As $d$ approaches $d_c^2$, the amplitudes $F_j^2$
increase drastically, whereas just slight increase of $F_j^1$ is observed with $d
\rightarrow d_c^2$, reflecting very weak response of the first facet on the critical
behavior of the second facet. Moreover, we find that correlations between fluctuations
on the different facets (measured by $F_j^{12}=\langle \Theta_j^1 \Theta_j^2 \rangle-
\langle \Theta_j^1\rangle \langle \Theta_j^2\rangle$ - Fig.~4, insets) are
practically zero compared to correlations within each facet (measured by $F_j^i$).
An analogous (but reversed) picture emerges close to the other cusp point. 
This behavior is consistent with the observed weak correlation between
fluctuations on different facets of Pt field-emitter tips \cite{fluct}.

The distinct character of fluctuations on different facets is also clear when analyzing
time series for $\Theta_j^1$ and $\Theta_j^2$, or the normalized probability
distributions $P(\Theta_j^i)$ (with $\int dx P(x)=1$). Fig.~5 shows $P(\Theta_j^i)$
obtained near the midpoints of the first and second bistable regions close to cusp
points. As $d \rightarrow d_c^1$ (Fig.~5a), the distributions of $\Theta_{\rm CO}^1$ and
$\Theta_{\rm O}^1$ become bimodal, reflecting a transition between reactive and inactive
states on the first facet induced by increased fluctuations.  However, the form of 
$\Theta_{\rm CO}^2$ and $\Theta_{\rm O}^2$ distributions remains effectively mono-modal.
Since the difference in $\Theta_j^2$ in the two stable states of bistability region \#1
on the second facet is very small (see coverages vs. $p_{\rm CO}$ in Fig.~1), the
transition between these states is practically invisible in the $\Theta_j^2$ time series
(Fig.~5(a)). Analogous (but reversed) behavior is observed as $d \rightarrow d_c^2$ 
(Fig.~5(b)). 

\subsection{Medium Coupling Regime}

To study what happens when the cusp point ($p_c^1$, $d_c^1$) nears ($p_c^2$, $d_c^2$) in
the ($p_{\rm CO}$, $d$)-plane, we consider next the medium-coupling regime when the
difference $|s_{\rm O}^1-s_{\rm O}^2|$ is small. We set $s_{\rm O}^1=0.9$,
$h=0.018$ , and analyze the $\Theta_j^i$ time series close to the {\it first} cusp point
($p_c^1 \approx 0.383$, $d_c^1 \approx 0.04$) of the bifurcation diagram
shown in Fig.~6(a). Specifically, we examine the midpoint of 
the larger bistability region \#2, close to the upper boundary of the inner
region \#1 (Fig. 6(b), $d=0.035$ and $p_{\rm CO} \approx 0.375$). Besides the
conventional fluctuation-induced transition between two states with low and high
$\Theta_{\rm CO}^1$ in bistability region \#1 (for
which $\Theta_{\rm CO}^2$ is always low, and thus distinct transitions in this quantity
are masked), the time series for $\Theta_{\rm CO}^1$ show the occurrence of additional
transition between two states with high $\Theta_{\rm CO}^1$. These derive from
bistability \#2, as is clear from the corresponding transitions in the time series for
$\Theta_j^2$. This produces a rather complicated form of the distributions for
$\Theta_{\rm CO}^1$, with additional peak splitting. Since the second cusp point
($p_c^2$, $d_c^2$) is close, corresponding fluctuations of the coverages in the states
of region \#2 on the first facet are sufficiently large to induce the additional
transition. Note that the time series shown on Fig.~6(b) are synchronized for both
facets due to the inter-facet CO diffusion.

\section{SUMMARY}

We have shown that the diffusion of adsorbed CO between two small adjacent 
facets in catalytic CO oxidation leads to the appearance of complex 
multistability on each facet. This replaces the conventional simpler bistability
with coexisting reactive and inactive stable steady states. The additional
stable steady states arise due to the difference in the sticking properties of the
facets, and the finite rate of inter-facet CO diffusion in our model (whereas
the CO mobility within each facet is assumed to be infinite). The topology of
bifurcation diagrams ($p_{\rm CO}$, $d$), as well as the effect of coverage
fluctuations, are shown to be qualitatively different in regimes with different
strengths of inter-facet coupling. In the weak-coupling regime the correlations between
the fluctuations near the cusp points terminating bistable regions are low. However,
in the medium-coupling regime, the transitions between the additional bistable
states on each facet can occur due to the communication between facets via 
CO diffusion.

\section*{ACKNOWLEDGEMENTS}

N.P. gratefully acknowledges the support of this research by 
Alexander von Humboldt Foundation and thanks the Ames Laboratory for warm
hospitality and providing with computational facilities. D.-J.L. and J.W.E. were
supported for this work by the Division of Chemical Sciences, USDOE, through Ames Laboratory
(operated by for the USDOE by Iowa State University under Contract No.~W-7405-Eng-82).

\newpage 

\section*{Figure captions} 

\noindent {\bf Fig.~1.} $p_{\rm CO}$-dependences of CO coverages for the facets with (a)
$s_{\rm O}^1=0.2$ and (b) $s_{\rm O}^2=1.0$ at $d=0.005$ and $h=0.01$. The insets show the
corresponding dependences for the isolated facets obtained using pair-approximation.\\[0.05in] 
{\bf Fig.~2.} Bifurcation diagrams ($p_{\rm CO}$, $d$) showing the bistable regions \#1 and \#2
calculated for different coupling
regimes at $s_{\rm O}^1=0.2$ and $s_{\rm O}^2=1.0$. \\[0.05in] 
{\bf Fig.~3.} Dependence of the critical parameters $h_1$ and $h_2$ on the $s_{\rm O}^1$ at
fixed $s_{\rm O}^2=1.0$. The insets show $p_{\rm CO}$-dependences of the steady-state CO coverages 
on the two facets at $s_{\rm O}^1=0$, $s_{\rm O}^2=1.0$ and $h=0.02$. (b) Bifurcation diagram calculated using
pair approximation at $s_{\rm O}^1=s_{\rm O}^2=1.0$ and $h=0.01$ (medium-coupling regime).
\#1 and \#2 indicate bistable region \#1 (where both "unsymmetric" steady states are stable) and
bistable region \#2, respectively.
The inset shows the corresponding $p_{\rm CO}$-dependence of $\Theta_{\rm CO}^1=\Theta_{\rm
CO}^2=\Theta_{\rm CO}$ obtained at $d=0.005$ with the stable and unstable states
indicated by black and gray colors respectively.\\[0.05in] 
{\bf Fig.~4.} Normalized amplitudes 
$F_j^i=\left[ N_i \langle \delta (\Theta_j^i)^2\rangle / \left(\langle
\Theta_j^i\rangle(1-\langle \Theta_j^i\rangle) 
\right)\right]^{1/2}$ 
and correlations $F_j^{12}=\langle \Theta_j^1 \Theta_j^2 \rangle-
\langle \Theta_j^1\rangle \langle \Theta_j^2\rangle$
of coverages fluctuations on weakly coupled facets for species $j=$~CO and O vs. $d$ 
for $p_{\rm CO}\approx 0.4096$ close to the cusp point of bistability \#1; 
$L=60$ and 700000 MC steps are taken for combined MC procedure.\\[0.05in]
{\bf Fig.~5.} Time series and distributions of $\Theta_j^i$ close to midpoint of 
(a) bistable region \#1, (b) bistable region \#2 ($h=0.005$, $s_{\rm O}^1=0.7$, $s_{\rm O}^2=1.0$).\\[0.05in]
{\bf Fig.~6.} (a) Bifurcation diagram calculated using combined MC scheme 
at $s_{\rm O}^1=0.9$, $s_{\rm O}^2=1.0$ and $h=0.018$ for the facets with medium coupling strength.
\#1 and \#2 indicate bistable regions \#1 and \#2, respectively. The inset shows
the stationary solutions $\Theta_{\rm CO}$ vs.
$p_{\rm CO}$ on each facet at $d=0.035$. (b) $\Theta_j^i$ distributions and
corresponding time series obtained at  $d=0.035$ and $p_{\rm CO}=0.375$ close to the
cusp point of bistability \#1 in the bifurcation diagram (a).\\[0.05in]









\end{document}